# Decarbonizing Mongolia's Energy Sector: A Techno-Economic Analysis of Hybrid Energy Solutions

Otgonpurev Nergui[1*], Dhammawit Paisiripas[2], Munkhsaikhan Jargalsaikhan[1]

***Abstract:*** *To achieve carbon neutrality and enhance energy security, Mongolia is exploring a transition toward hybrid energy solutions integrating small modular reactors (SMRs) and renewable energy sources. This study assesses the feasibility of a grid-connected hybrid energy system that combines coal, solar photovoltaic (PV), wind turbines, battery energy storage systems (BESS), and SMRs to reduce greenhouse gas emissions and improve energy reliability. Using the HOMER software, multiple energy mix scenarios are evaluated based on net present cost (NPC), levelized cost of energy (LCOE), and $CO_2$ emissions. The optimal scenario (Scenario 4) demonstrates a significant reduction in $CO_2$ emissions, projecting a decrease to 6,755,129 metric tons by 2030 compared to the current system's 10,860,025 metric tons. Additionally, this scenario achieves an NPC of \$31.5 billion and an LCOE of \$0.0801/kWh, demonstrating both economic viability and environmental benefits. The findings highlight the potential of SMRs and renewables in Mongolia's energy transition, providing valuable insights for policymakers seeking a sustainable and decarbonized energy future.*

[1*] Otgonpurev Nergui, Nuclear Research Center, National University of Mongolia, otgonpurev@num.edu.mn
[2] Paisiripas Dhammawit, The Electricity Generating Authority of Thailand (EGAT), dhammawit.pa@egat.co.th
[1] Munkhsaikhan Jargalsaikhan , Nuclear Research Center, National University of Mongolia, j.munkhsaikhan@num.edu.mn

## 1. INTRODUCTION

Mongolia is a landlocked country in East Asia, bordered by Russia to the north and China to the south. Most of the country is hot in the summer and extremely cold in the winter, with January averages dropping as low as −30 °C (−22 °F). As a result, both electricity and heat supply play a crucial role in the country's energy infrastructure. **[1]**

The majority of Mongolia's energy is derived from coal, with a smaller portion coming from crude oil exports, mainly to China. According to the World Energy 2020 statistics, 84% of global energy still relies on fossil fuels, while renewable energy accounts for only 11% of global primary energy consumption. The continued reliance on fossil fuels significantly contributes to greenhouse gas (GHG) emissions, exacerbating climate change and environmental pollution. **[2]**

In contrast, renewable energy sources offer a cleaner alternative, producing substantially lower emissions. Mongolia's energy sector faces several challenges, including its vast and sparsely populated landscape, outdated energy infrastructure, and high dependence on coal-based district heating systems. Many rural areas still rely on inefficient diesel generators and coal-fired boilers, which often operate only a few hours per day due to fuel shortages and financial constraints.

Rapid economic growth and urbanization have led to a significant increase in energy demand, particularly in Ulaanbaatar, Mongolia's capital, where over 1.6 million people reside. This growing demand has placed considerable pressure on the country's aging energy infrastructure. **[3]** Mongolia is actively pursuing energy diversification to meet its growing electricity demand and transition toward a more sustainable energy future. State Policy on Energy (2015–2030) seeks to ensure a stable and reliable power supply while positioning the country as a future energy exporter. **[4]** As of 2023, Mongolia's electricity generation predominantly relied on combined heat and power (CHP) plants, accounting for 90.9% (1264MW) of total production.

In contrast, renewable energy sources contributed 8.5% (wind turbine 155MW, solar PV 115.7MW), while hydropower and diesel generators supplied 0.6% (26.4 MW) and 0.01%,(8.55 MW) respectively. However, diesel generators and hydroelectric power plants only produce a limited amount of electricity, so this study will not consider electricity generation from these sources. Figure 1 illustrates Mongolia's production of energy by source. **[5]**

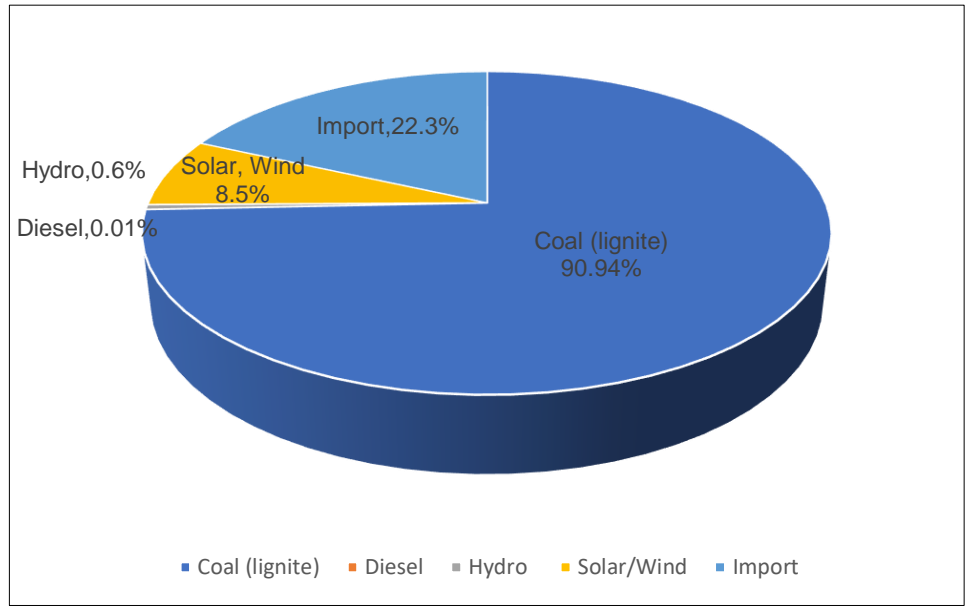

Figure 1. Mongolia Electricity Generation Mix 2023,(%)

Source: International Energy Agency https://www.iea.org/countries/mongolia/energy-mix

(accessed February 17, 2025).

Electricity generation remains one of the largest sources of $CO_2$ emissions in Mongolia. In 2023, total electricity production reached 8,528.27 million kWh, reflecting a 4.3% increase from the previous year. The heavy reliance on coal-fired power plants has led to severe air pollution, posing significant public health risks. To mitigate these challenges, the government of Mongolia has established significant renewable energy objectives, intending to elevate the proportion of renewables in the national energy mix to 30% by 2030.

This study investigates the feasibility of a grid-connected hybrid energy system that integrates coal, solar PV, wind turbines, battery energy storage systems (BESS), and small modular reactors (SMRs) to meet Mongolia's energy needs. The HOMER Pro software is used in this paper for optimization and techno-economic analysis.

The findings provide insights into the potential for diversifying Mongolia's energy sources and offer valuable recommendations for policymakers to support a more sustainable and decarbonized energy future.

## 2. METHODOLOGY

To evaluate the performance of a hybrid renewable energy system, this study employed the HOMER (Hybrid Optimization of Multiple Energy Resources) software, a widely used simulation tool for modeling off-grid and grid-connected power systems. HOMER facilitates optimization by simulating different system configurations based on user-defined technical and economic parameters **[6].**

**Mathematical modeling of hybrid system**

1. *PV array power output model*

The solar energy output modules can be calculated according to the following equation **[7].**

$$P_{pv} = f_{pv} Y_{pv} \frac{G_T}{G_{T,STC}} [1 + C_p (T - T_{STC})] \quad (1)$$

where $f_{PV}$ is PV derating factor; $Y_{PV}$ is the rated capacity of the PV array under standard test condition (W); GT is the global solar radiation incident on the surface of the PV array (kW/m$^2$); and $G_T$, $S_{TC}$ is the standard radiation at standard test condition (1kW/m$^2$). Cp is the temperature coefficient of power (%/C), $T_C$ is the PV module temperature in real-time (°C), and $T_{STC}$ is the PV module temperature under standard test conditions (25°C).

In the study, we will use Canadian Solar MaxPower CS6X-325P **[8]** to simulate photovoltaic panel components in HOMER.

2. *Wind turbine model*

HOMER calculates the power output of the wind turbine in each time step using a three-step process. First, HOMER calculates the wind speed at the hub height of the wind turbine. Then it calculates how much power the wind turbine produces at that wind speed at standard air density. Finally, the adjusts that power output value for the actual air density **[9]**

$$V_{HUB} = V_A \frac{\ln(\frac{H_{HUB}}{H_0})}{\ln(\frac{H_A}{H_0})} \quad (2)$$

Power curves typically specify wind turbine performance under conditions od standard temperature and pressure. To adjust actual conditions, power output value for the actual air density is calculated as the following equation:

$$P_{WTG} = \left(\frac{\rho}{\rho_0}\right) P_{WTG,STP} \quad (3)$$

where P$_{WTG}$ = output power (kW), P$_{WTG,STP}$ = the wind turbine power output at standard temperature and pressure [kW], $\rho$= real air density (kg/m³), and $\rho_0$= air density at standard pressure and temperature (1.225 kg/m3).

Mongolia has 3 wind turbines named Salkhit (50 MW), Tsetsii (50 MW), and Shand-Salkhit (55 MW), with a total capacity of 155 MW. The Enercon E-82 E2 wind power model will be used in the project. **[10]**

### 3. Battery model

The Tsengeg 0.6 MW and Songino 80 MW battery energy storage systems (BESS) are set to be commissioned in Mongolia in 2023. This study applies a generic Li-Ion battery package with a capacity of 1 MWh and a nominal voltage of 600 V. The following equation describes the storage capacity of a battery **[11]**

$$C_{wh} = E_L \cdot AD \cdot \eta_{inv} \cdot \eta_{BAT} \cdot DOD \quad (4)$$

where, $C_{wh}$ is battery capacity (kW), $E_L$ is total load demand (kWh/day), $AD$ is daily autonomy, $DOD$ is depth of discharge; $\eta_{inv}$ is inverter efficiency, and $\eta_{BAT}$ is battery efficiencies.

### 4. Converter

The converter facilitates energy flow between AC and DC components within the hybrid system. It was sized based on peak load requirements and system design specifications **[12]**. A general big, free converter model was used in this study.

### 5. Small Modular Reactor(i- SMRs)

The SMR was modeled after a generic large generator set component because HOMER does not include a nuclear power plant module. Reactor design, location, and regulatory requirements are only a few of the variables that affect the capital cost of compact modular reactors. The late at night capital cost is typically estimated to be between $3000 and $10,000 per kW **[13, 14].** There are no carbon gas emissions when SMR is in operation. However, if fossil fuels are utilized in the mining or processing of uranium ore, greenhouse gas emissions that result from those activities are not taken into consideration in this analysis.

6. *CHPPS (Coal-fired power plant)*

Because the HOMER program lacks a coal-fired power plant model in its modeling capabilities, the CHPPs were built using generic large generator components.

As of today, the average price of coal with a calorific value of 5,500 kcal/kg is 452,306 MNT. Using the exchange rate of 1 USD = 3,459 MNT (as of February 14, 2025), the price for 5,500 kcal/kg of coal is approximately $0.15 per kg **[15]**.

7. *Grid*

As of November 15, 2024, Mongolia implemented a tiered electricity tariff system to promote energy efficiency and ensure the financial sustainability of its energy sector **[16]**. A three-tier tariff based on consumption is to be introduced. The average household tariff was previously MNT 140, but under the revised tariff, consumption up to the first 150 kWh in a month will be charged at MNT 175, consumption between 150-300 kWh at MNT256, while consumption exceeding 300 kWh will be MNT285. Approximately 85% of Mongolia's electricity is generated from coal-fired power plants. Given the high reliance on coal, Mongolia's electricity consumption is associated with substantial $CO_2$ emissions. For instance, consuming 1,000 kWh (1 MWh) of electricity would result in approximately 582 g/kwh of $CO_2$ emissions. When evaluating energy projects or consumption in Mongolia, it's crucial to factor in this high GEF to accurately assess environmental impacts and explore opportunities for integrating more renewable energy sources to reduce the carbon footprint **[17]**.

8. *Economic Model*

A. NPC-Net Present Cost

The Net present cost (or life cycle cost) of hybrid system is the present value of all the costs of installing and operating the systems over the project lifetime, minus the present value of all the revenues that it earns over project lifetime.

$$NPC = \frac{TAC \cdot (1+i)^n - 1}{i(1+i)^n} \qquad (5)$$

where, $TAC$ is total annualized costs ($/year), i is interest rate (%), and n is project lifetime (years).

B. LCOE- Levelized Cost of Energy

The average cost per kWh of power generated by the system is known as the levelized cost of energy, or LCOE. The LCOE is calculated by adding the project's annual costs ($/year) to the total electrical load (kWh/year) produced by the energy system. This equation can be used to calculate it **[18]**.

$$LCOE = \frac{TAC}{E_{Load}} \qquad (6)$$

where, $TAC$ is total annualized costs ($/year), and $E_{Load}$ is the total electrical load (kWh/year).

9. *Emissions*

There were six pollutants estimated in research such as carbon dioxide ($CO_2$), carbon monoxide (CO), unburned hydrocarbons (UHC), particulate matter (PM), sulfur dioxide ($SO_2$), and nitrogen oxides (NOx). Among these, $CO_2$ is used as representing a measure of emission. The annual $CO_2$ emissions from a generator are calculated from the below equation **[19]**.

$$E_{gen} = E_{factor} \cdot T_{fuel} \qquad (8)$$

where, $E_{gen}$ refers to the emissions of $CO_2$ by generators, $E_{factor}$ is the emissions factor amount of $CO_2$ emitted per unit fuel consumption, and $T_{fuel}$ is the total fuel consumption by the generators.

10. *Renewable Fraction*

The percentage of energy generated from renewable resources that is supplied to the load is known as the renewable fraction. It is shown as a percentage. The following formula was used to get the renewable fraction **[14, 20]**.

$$RF = 1 - \left(\frac{E_{g,non-ren}}{E_{Load}}\right) \qquad (9)$$

where, $E_{g,on-ren}$ is the non-renewable electrical production (kWh/year), and $E_{Load}$ is the total electrical load (kWh/year).

## 3. CASE STUDY
*3.1 Renewable Energy Potential of Mongolia*

Mongolia possesses significant potential for renewable energy development. The country enjoys 270–300 sunny days per year, with an estimated 2,250–3,300 daylight hours annually. The Gobi Desert, which spans 5,542 km² and shares a border with China, is among the world's top regions for solar energy generation, with an average solar radiation of 5.4 kWh/m²/day. Additionally, the desert features abundant wind resources, with wind speeds exceeding 9.0 m/s at 80 meters and an annual wind duration of 4,000–4,500 hours, making it an ideal location for large-scale wind energy projects. **[21]** This study is based on the solar radiation and the wind speed data of Mongolia obtained from the NASA's data of predicted global resources for energy. **[22]**

The solar radiation across the area of interest is calculated to make the best use of the available daylight energy. The average daily solar radiation for the year is 5.71 kWh/m²/day, varying from 3.74 kWh/m²/day in January to 7.82 kWh/m²/day in May, as shown in Figure 2.

The mean annually clearness index is around 0.5, calculated from monthly averages of worldwide horizontal radiation for 22 years (July 1983–June 2005). **[23]**

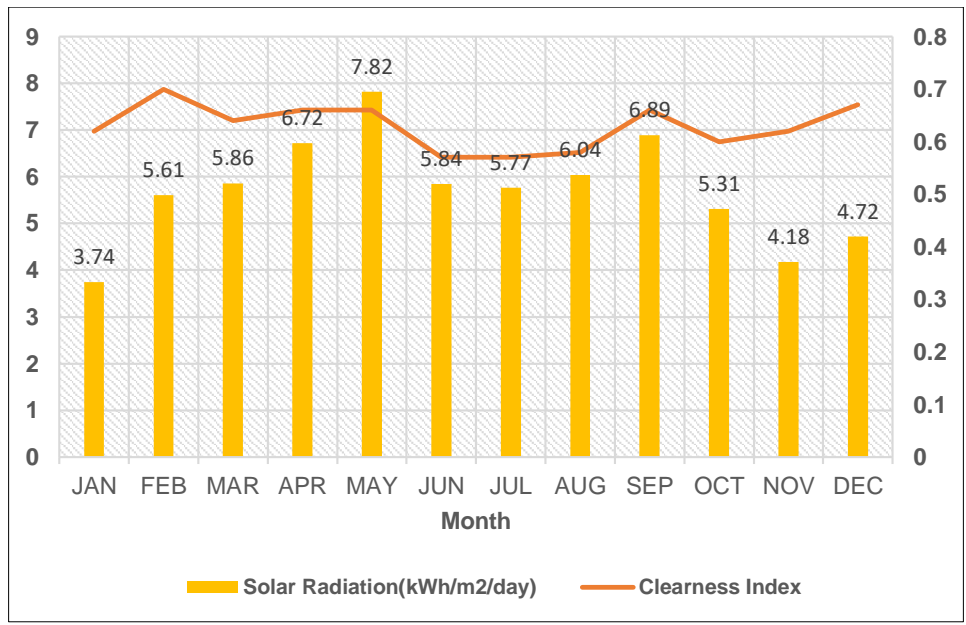

Figure 2 illustrates the average monthly sunlight radiation along with the clearness index.

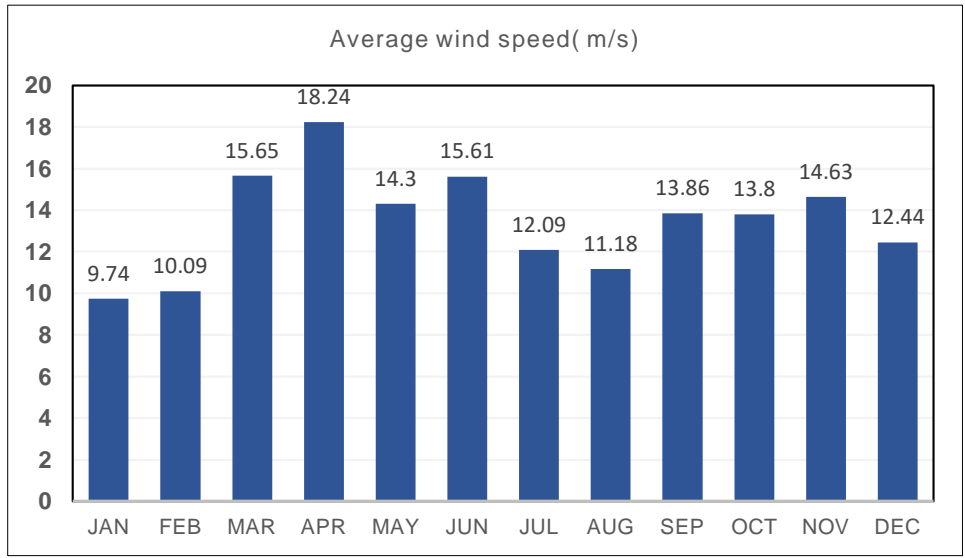

Figure 3. Annual mean wind speed in Mongolia

The annual average wind speed is 18.48 m/s at an elevation of wind speed at 50m range. The maximum average wind speed is 18.24 m/s in April while the minimum average wind speed is 9.74 m/s in January **[24]**. Monthly wind speed of 2022 in Mongolia shown in Figure 3.

3.2 *Load Profile*

Mongolia's power production in 2023 was 8528.7 million kWh, and imported electricity was 2447.6 million kWh, with a total electricity consumption of 10,975.9 million kWh during the study period. In the winter of 2023–2024, the Central Power System had a high load of 1636 MW on 21 December 2023 **[25].**

Mongolia's electric load profile for 2023 was obtained for simulation purposes. The median everyday demand for load in Mongolia was 3,0175,000 kWh/day with a 1,636 MW peak as shown in Figure 4. This was used as the primary load input needed for the analysis with HOMER.

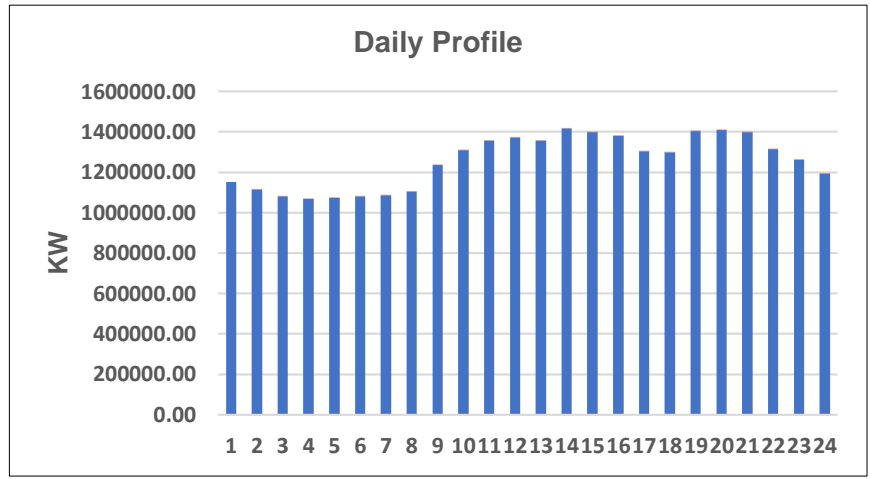

Figure 4. Estimated load profile of Mongolia

This paper aims to use Mongolia's projected electricity demand in 2030 more practically in future renewable energy projects. In line with the State Policy on Energy (2015-2030), which aims to increase renewable energy to constitute 20% of Mongolia's total power capacity by 2020 and 30% by 2030. Between 2023 and 2030, Mongolia's net electrical consumption is projected to expand at a mean annual rate of approximately 7-8%. This projection is based on recent data indicating that the country's electricity consumption has been increasing by about 7-8% annually, while installed energy capacity has been growing by 5-6% per year.

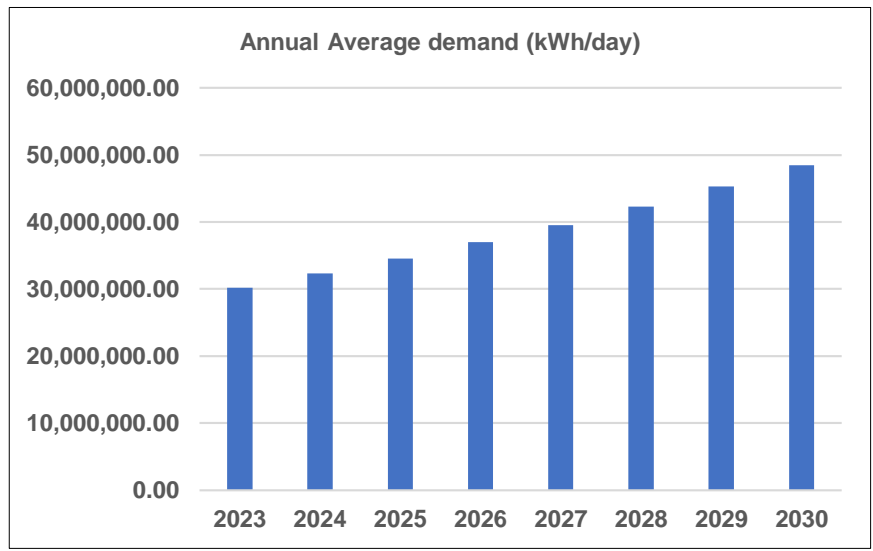

Figure 5. The projected load demand of Mongolia from 2023 to 2030

Figure 5 illustrates the anticipated load requirement. After scaling, the Mongolia load profile will be projected to 48,454,456 kWh/day with a peak load of 2,628 MW, as shown in Table1.

Table 1. Scaled load demand of Mongolia for the year 2030.

| Metric | Baseline | Scaled |
|---|---|---|
| Average (kWh/d) | 30,175,000,00 | 48,454,456,05 |
| Average (kW) | 1,257,291.67 | 2,018,935.67 |
| Peak (MW) | 1,636,149.45 | 2,627,298,48 |
| Load factor | 0.77 | 0.77 |

## 4. RESULTS AND DISCUSSION

### 4.1 Simulation Results

In this study, various scenarios based on techno-enviro-economic perspectives for hybrid systems were simulated and compared regarding supply load demand.

#### 4.1.1 Scenario 1: Current system ( CHPPs/Grid/PV/WT/BESS)

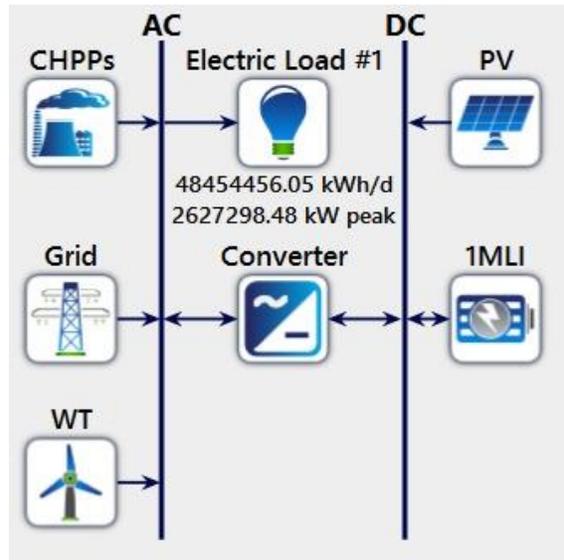

Figure 6. Schematic diagram of reference scenerio.

The reference scenario, depicted in Figure 6, is modeled to include all energy system components across Mongolia. As of 2023, Mongolia's electricity generation predominantly relied on combined heat and power (CHP) plants, accounting for 90.9% (1264MW) of total production.

In contrast, renewable energy sources contributed 8.5% (wind turbine 155MW, solar PV 115.7MW), while hydropower and diesel generators supplied 0.6% (26.4 MW) and 0.01%,(8.55 MW) respectively. In addition, Mongolia's electricity production reached 8,528.7 million kWh, with an additional 2,447.6 million kWh imported **[26].**

The simulation results for 2030 show that system electricity production is 17,656 million kWh, while imported electricity reaches 5,869 million kWh.

From the scenario 1, the total NPC is 40.1 billion USD and LCOE is 0.102 USD/kWh, respectively, and this system generates 10,860,025 tonnes/year of $CO_2$ annually.

### 4.1.2 Scenario 2: Current system+New CHPPs

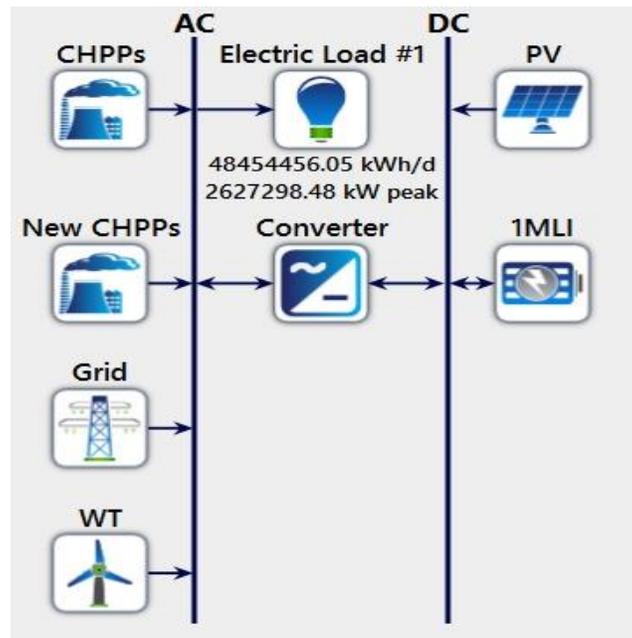

Figure 7. Scenario2: Current system+New CHPPs

As outlined in the "State Energy Policy 2015-2030" , the most anticipated energy development within the framework of the goal of doubling the installed capacity of Mongolia's energy resources is the Ulaanbaatar Thermal Power Plant No. 5 project. **[3, 27].**

Under this policy, scenario 2 will be modeled, where the current energy system is integrated with the 450 MW generated by the New CHP, as illustrated in Figure 7.

The total NPC and LCOE of the combination have been estimated at 40.4 billion USD and 0.103

USD/kWh, respectively, and this system generates 10,991,899 tonnes/year of $CO_2$ annually.

*4.1.3   Scenario 3: Current system+ i-SMR*

In this scenario, electricity from the current system is integrated with 680 MW of electricity produced from the i-SMRs, as shown in Figure 8, assuming four modules of i-SMR are deployed on the Mongolia.

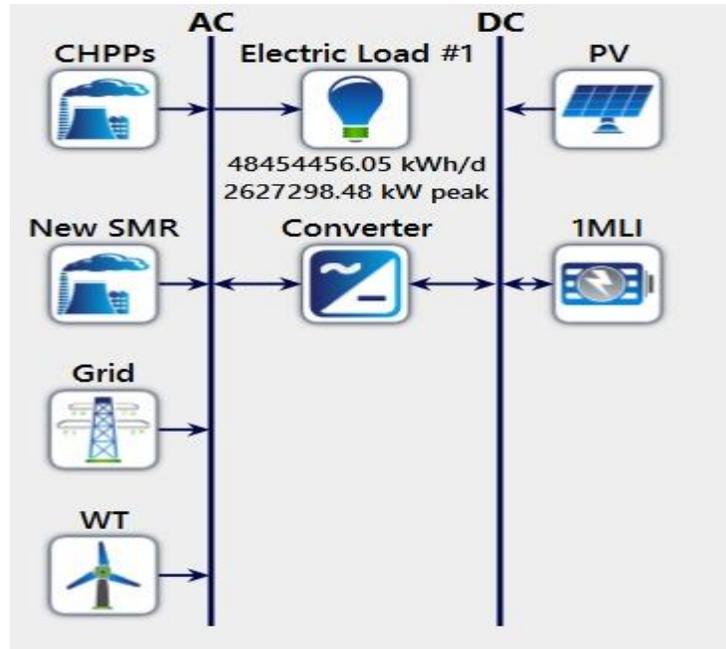

Figure 8. Scenario3: Current system+ i-SMR

This simulation aligns with Mongolia's long-term development strategy, the New Recovery Policy (2041-2050), which seeks to harness nuclear energy. **[28]** The simulation result showed the total NPC is 31.3 billion USD, the LCOE is 0.0795 USD/kWh, and the $CO_2$ generated from this system is 7,330,300 tonnes per year.

### 4.1.4 Scenario 4. Current+iSMR+ 300MW New WT

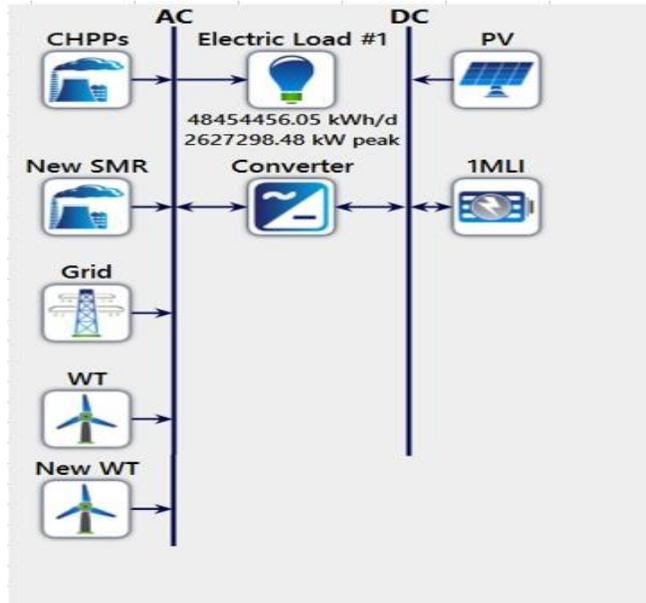

Figure 9. Scenario 4. Current+iSMR+ 300 MW New WT

Currently, 3 wind turbines operate with a total capacity of 155 MW in Mongolia. In that case, it is simulated by assuming that wind turbines with a capacity of 300 MW each are integrated into the system, as shown in Figure 9. The total NPC is 31.5 billion USD, the LCOE is 0.0801 USD/kWh, and the $CO_2$ generated from the system is 6,755,129 tonnes per year.

### 4.1.5 Scenario 5. Current +iSMR+ 300MW new PV

This scenario illustrates the combination of coal, grid, nuclear, and renewable energy sources. Figure10 provides a schematic of a grid-connected hybrid energy system that includes a current system that integrates 680 MW of i-SMR and 300 MW of new PV.

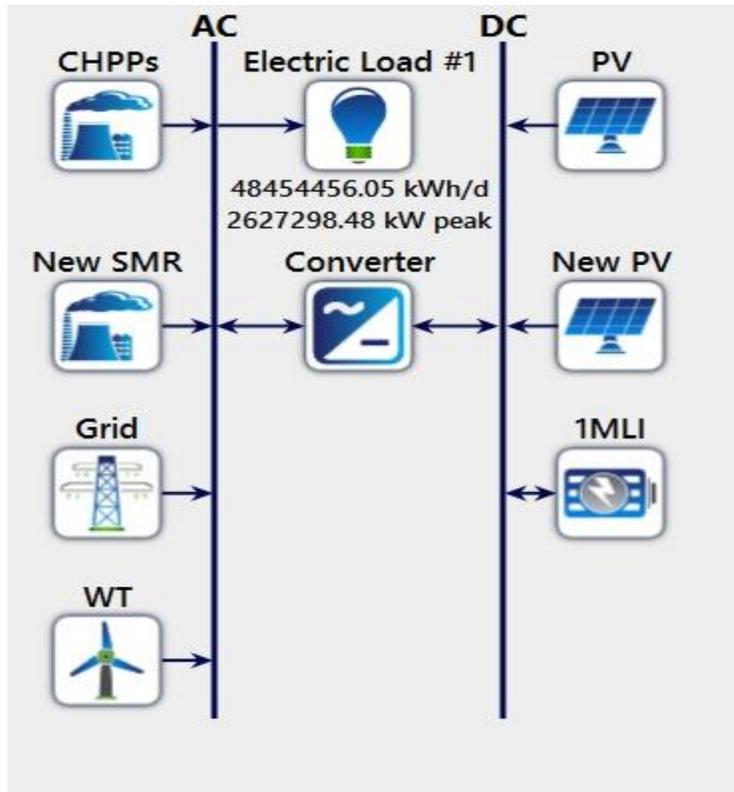

Figure 10. Scenario 5. Current +iSMR+ 300MW new PV

The total NPC is 32.3 billion USD, the LCOE is 0.0821 USD/kWh, and the $CO_2$ generated from the system is 6,986,719 tonnes per year.

5. *Comparison Analysis of Hybrid Energy Systems*

To address Mongolia's increasing energy demand, this study aims to develop a hybrid energy system that diversifies energy sources and enhances sustainability and reliability. This section explores the techno-economic and environmental aspects of hybrid energy systems. The simulation results of the hybrid energy systems are organized from the most feasible to the least attractive. Table 2 presents the results ranked by NPC, R.F. LCOE, and $CO_2$ emission.

Table 2. Simulation results of the hybrid renewable energy systems ranked by NPC

| Rank | Scenario | Hybrid system | NPC (Billion $) | LCOE ($/kWh) | R.F(%) | $CO_2$ emission (tonnes/year) |
|---|---|---|---|---|---|---|
| 1 | 4 | Current system/i-SMR/300MW new WT | 31.5 | 0.0801 | 8.88 | 6,755,129 |
| 2 | 3 | Current system/ i-SMR | 31.3 | 0.0795 | 3.85 | 7,330,300 |
| 3 | 5 | Current system/i-SMR/300MW new PV | 32.3 | 0.0821 | 6.99 | 6,986,719 |
| 4 | 2 | Current system/new CHPPs | 40.4 | 0.103 | 3.87 | 10,991,899 |
| Ref. case | 1 | Current system (CHPPs/Grid/PV/WT/BESS) | 40.1 | 0.102 | 3.84 | 10,860,025 |

The results show that the most favorable scenario (4) from an environmental perspective is an NPC of $31.5 billion and an LCOE of $0.0801/kWh. The most favorable scenario from an economic perspective is scenario (3), with an NPC of $31.3 billion and an LCOE of $0.0795/kWh. The least favourable option was the isolated scenario (2), with an NPC value of 40.4 billion USD and LCOE at 0.103 $/kWh, respectively.

From an environmental perspective, scenario (4) reduces greenhouse gas emissions, producing 6,755,129 tonness of $CO_2$ annually, compared to the current system, which emits 10,860,025 tonnes of $CO_2$ annually. Minimizing greenhouse gas emissions is a crucial factor in electricity generation systems. In Mongolia, thermal power plants account for the majority of carbon dioxide emissions. This study will consider $CO_2$ emissions from the thermal power plant. As of 2023, a total of 11.36 million tons of carbon dioxide were emitted from thermal power plants. According to our study's optimal simulation (Scenario 4), CO2 emissions are projected to decrease to 6,755,129 metric tons by 2030.

## 6. Sensitivity Analysis of the Optimal Result

Sensitivity analysis assists in evaluating the impact of uncertainty or variable changes, such as variations in capital cost, discount rate, or electric load demand. The sensitivity analysis for the system's optimal scenario 4 is carried out in this section.

### 6.1.1 Effect of Electric Load Variations

A number of factors, including growth in the population, economic expansion, and climatic change, are taken into account when estimating the demand for power load. To determine the effect of electrical load demand on the performance of the ideal hybrid energy system, a sensitivity analysis must be carried out. In this section, the NPC and LCOE of the optimal hybrid energy system have been calculated by varying the electricity forecasted load demand from -10% to +10% with a 5 % interval. Figure 11 illustrates the result of sensitivity to the electricity load demand.

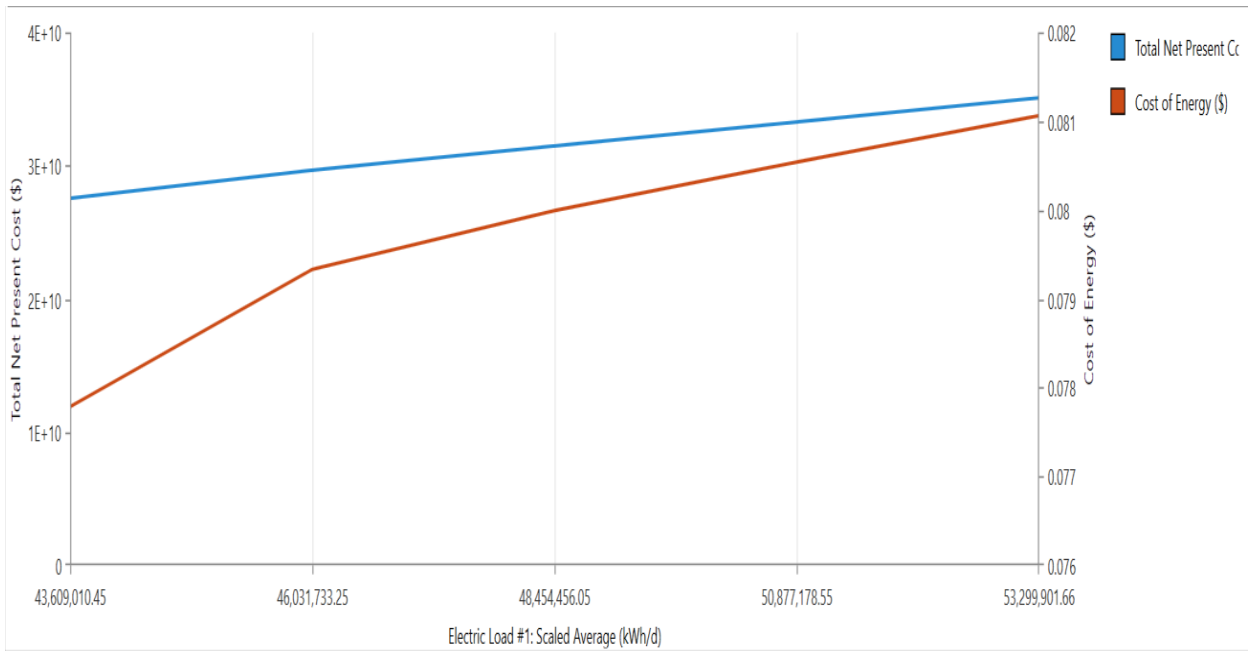

Figure 11. Effect of Electric Load Variations on NPC and LCOE.

### 6.1.2 *Effect of Discount Rate Variations*

The discount rate is a vital economic factor when assessing project viability. In this section, the optimal hybrid energy system's NPC and LCOE are analyzed by varying the nominal discount rate between 6%, 8%, 10%, 12%, and 14%. Figure 12 depicts the impact of changing nominal discount rates on NPC and LCOE.

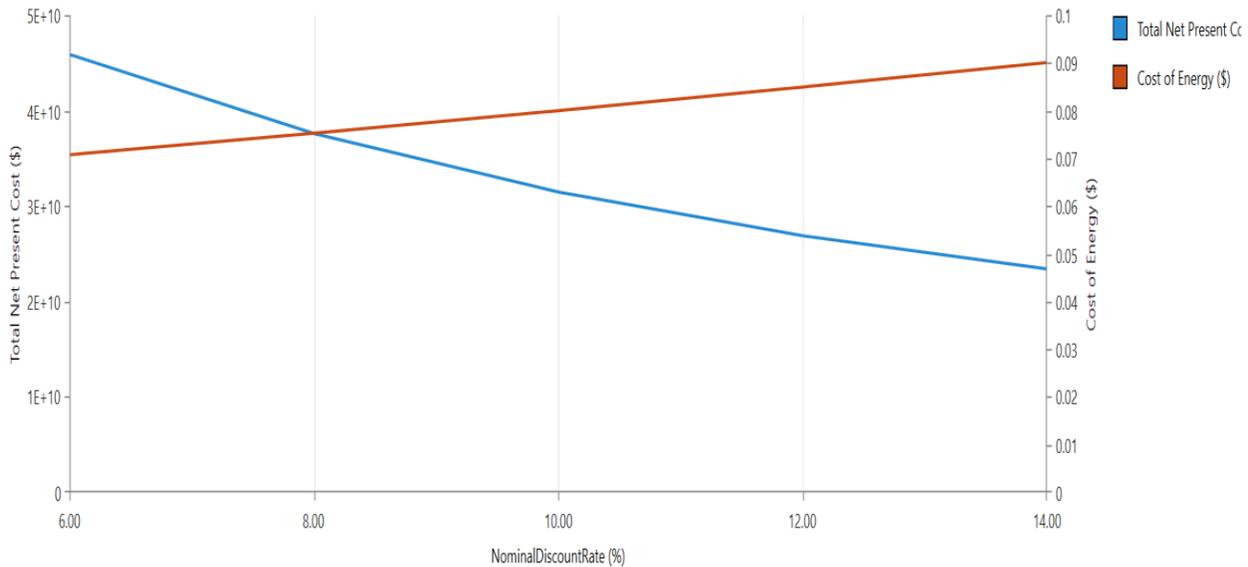

Figure 12. Effect of Discount Rate Variations on NPC and LCOE

### 6.1.3 *Effect of Capital Cost Variations*

The variation of the capital costs of different components of the scenario 4 system on the total NPC and LCOE is observed in Figure 13 and Figure 14. The capital cost of each component varies between 0.5 and 1.5. It is evident from both numbers that SMR's capital cost has the biggest impact on NPC and LCOE. Considering the highest cost element of the hybrid energy system is the capital cost of SMR.

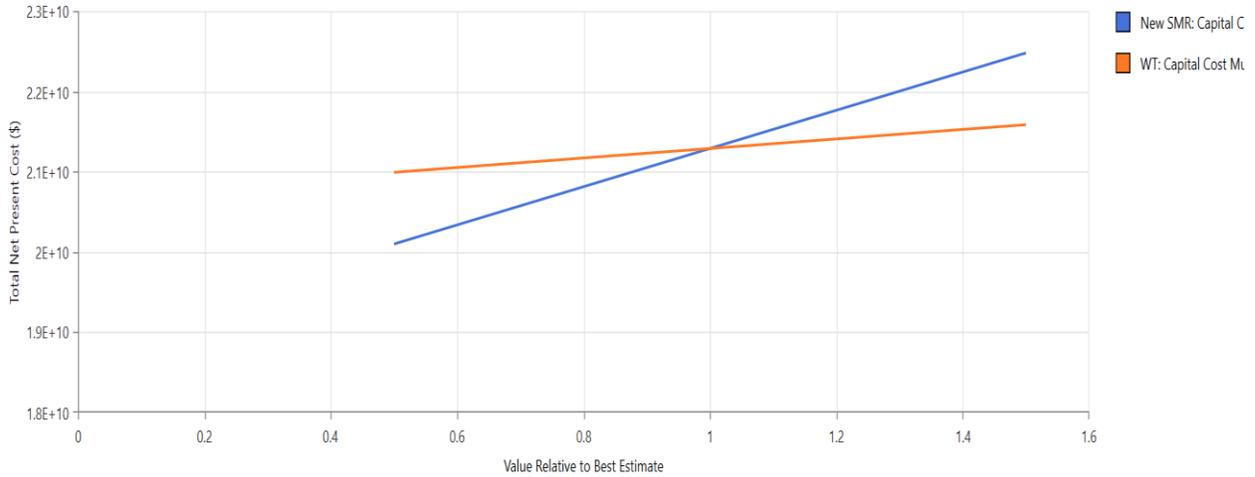

Figure 13. Effect of capital cost variations of NPC

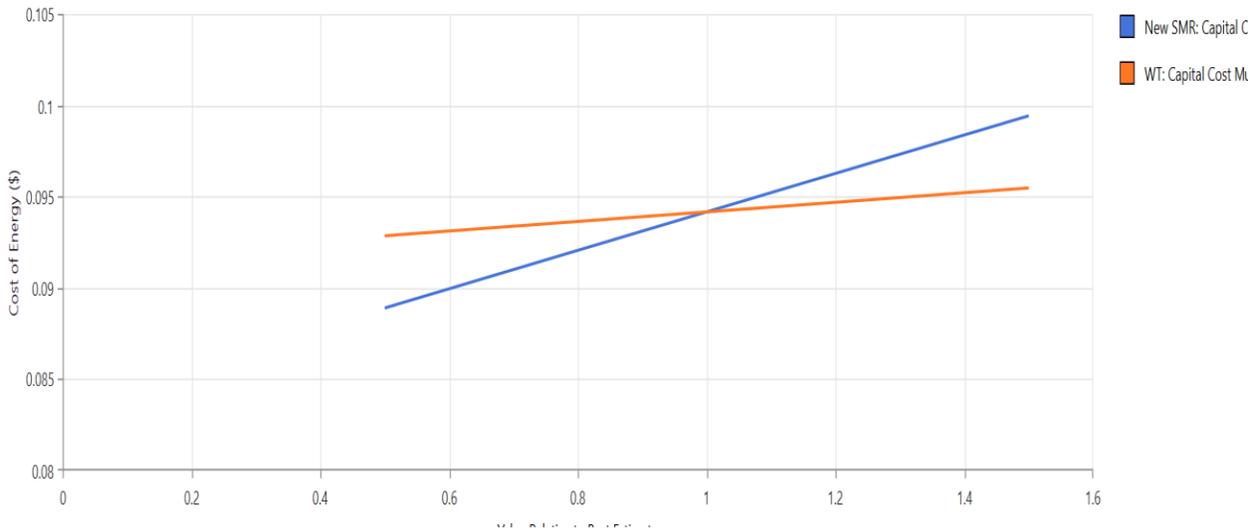

Figure 14. Effect of Capital Cost Variations on LCOE.

### 6.1.4 Effect of Electricity Price Variations

Electricity price fluctuations result from a complex interaction of factors such as supply and demand dynamics, fuel costs, regulatory policies, and seasonal variations. This section analyzes the NPC and LCOE, calculated by adjusting the electricity price from 0.08 to 0.12$/kWh.

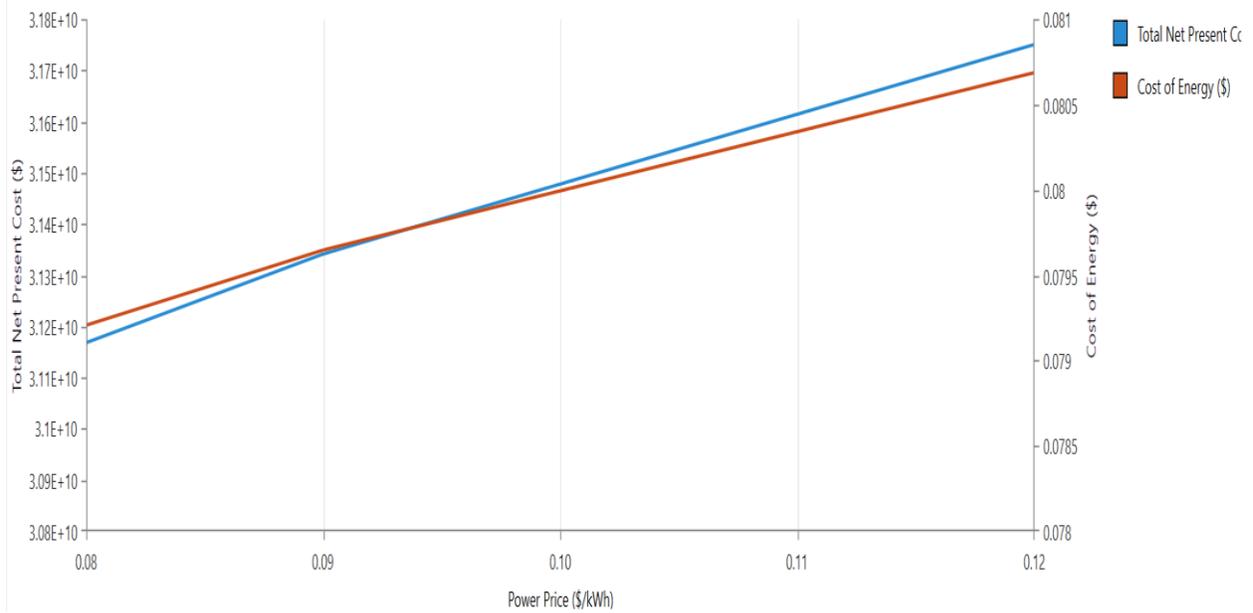

Figure 15. Effect of electricity price variations on NPC and LCOE .

## 7. CONCLUSION

This study evaluates the feasibility of integrating small modular reactors (SMRs) with renewable energy sources to reduce greenhouse gas emissions and enhance energy security in Mongolia. The study compares the economic, technical, and environmental impacts of different hybrid energy systems by simulating multiple energy mix scenarios using the HOMER software.

In addition to two environment indicators the renewable fraction and CO2 emissions two economic indicators NPC and LCOE were taken into account in this assessment. Four scenarios of hybrid systems were analyzed based on technical, environmental, and economic perspectives. Then sensitivity analyses were carried out for the optimal scenario. The results indicate that integrating SMRs with wind and solar energy significantly reduces Mongolia's reliance on coal-fired power plants while maintaining cost-effectiveness.

Among the evaluated scenarios, the hybrid system integrating SMRs with 680 MW and 300 MW of new wind power (Scenario 4) emerged as the most favorable option. It achieved the lowest net present cost (NPC) of 31.5 billion USD and a levelized cost of energy (LCOE) of 0.0801 USD/kWh. The key advantages of this optimal setup include:

1. LCOE Reduction: Scenario 4 achieves an LCOE of 0.0801 USD/kWh, representing a 21.5% decrease compared to Scenario 1 (0.102 USD/kWh), enhancing economic feasibility.
2. $CO_2$ Emissions Reduction: Scenario 4 reduces $CO_2$ emissions by 37.8%, lowering annual emissions from 10,860,025 tonnes in Scenario 1 to 6,755,129 tonnes, demonstrating significant environmental benefits.
3. NPC Reduction: With an NPC of 31.5 billion USD, Scenario 4 is 21.4% more cost-effective than Scenario 1 (40.1 billion USD), reinforcing its financial viability.

This analysis highlights Scenario 4 as the most economically and environmentally advantageous option. The findings confirm that hybrid energy solutions incorporating SMRs and renewables can be crucial in Mongolia's energy transition, supporting national decarbonization goals while ensuring a reliable and stable power supply. As Mongolia moves toward increasing its share of renewable energy, policymakers should consider the deployment of advanced nuclear technologies, such as SMRs, alongside wind and solar energy to achieve a sustainable, low-carbon energy future. Further research should explore site-specific implementation strategies, regulatory frameworks, and financial mechanisms to facilitate the large-scale adoption of hybrid energy systems in Mongolia.

*Funding:* No funding was received for this manuscript
*Data Availability Statement:* Data are contained within the article.
*Conflict of Interest:* The authors declare no conflicts of interest.